\documentclass[11pt]{article}
\usepackage{moriond2000a,epsfig}

\def\Journal#1#2#3#4{{#1} {\bf #2}, #3 (#4)}

\def\be{\begin{equation}}
\def\ee{\end{equation}}
\def\sm{M_{\odot}}

\def\macho{{\sc Macho}}
\def\eros{{\sc Eros}}

\newcommand{\la}{\;\raisebox{-.8ex}{$\buildrel{\textstyle<}\over\sim$}\;}

\begin{document}

\vspace*{4cm}
\title{MACHOs and the clouds of uncertainty}

\author{ Eamonn Kerins }

\address{Theoretical Physics, University of Oxford, \\ 1 Keble Road,
Oxford OX1 3NP, UK}

\maketitle

%%%%%%%% authors photo (option) %%%%%%%%%%%%%%%%%%%%% 
\begin{figure}[h]
\begin{center}
\psfig{figure=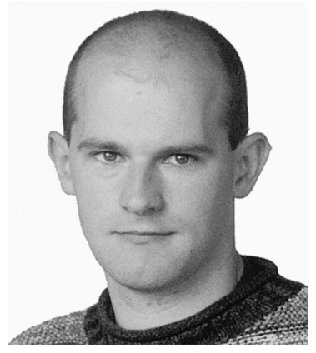}
\end{center}
\end{figure}

\abstracts{
I review proposals for explaining the current gravitational
microlensing results from the \eros \ and \macho \ surveys towards the
Magellanic Clouds. Solutions involving massive compact halo objects
(MACHOs), both baryonic and non-baryonic, as well as solutions that do 
not require MACHOs, are discussed. Whilst the existence and nature of MACHOs
remains to be established, the prospects for achieving this over the
next few years are good.
}

\section{Microlensing and MACHOs --- where we now stand}

\subsection{The \eros \ and \macho \ surveys}

\eros \ and \macho \ \cite{micrefs}, have been monitoring millions of
stars in the Large and Small Magellanic Clouds (LMC and SMC) on an
almost nightly basis since 1992, enabling them to search for MACHOs
with masses from around $10^{-4}~\sm$ up to several Solar
masses. \eros \ also undertook a survey of a smaller number of LMC
stars with a sampling of 30 mins, extending its sensitivity down to
$10^{-8}~\sm$. Other experiments are now also targeting the Magellanic
Clouds but have yet to publish full results, so in this review we
shall concentrate on the \eros
\ and \macho \ surveys. Details of these and the other surveys, and of
the principles of microlensing, are described elsewhere in these
proceedings.

The \macho \ experiment has published around fifteen
candidates towards the LMC and two candidates towards the SMC. One of
the LMC candidates, MACHO~LMC-9 \cite{ben96}, appears to be a
caustic-crossing event due to a binary lens, as is the SMC candidate,
98-SMC-1 \cite{alc00}. \eros \ has three candidate events towards the
LMC and two towards the SMC (including 98-SMC-1). Of the two targets,
the LMC has been the monitored the most intensively.
It is worth noting that there are three strong arguments against the
candidate microlensing events being instead some hitherto unknown
population of variable stars. Firstly, the impact parameter
distributions are consistent with microlensing expectation. Secondly,
the positions of the source stars on the HR diagram show no obvious
clustering. Thirdly, in the case of the microlensing experiments
looking towards the Galactic Bulge, the optical depth deduced only
from clump giant sources is consistent with that inferred from all the 
events implying that, at least along this line of sight, contamination 
levels are small.

\begin{figure}
\epsfig{file=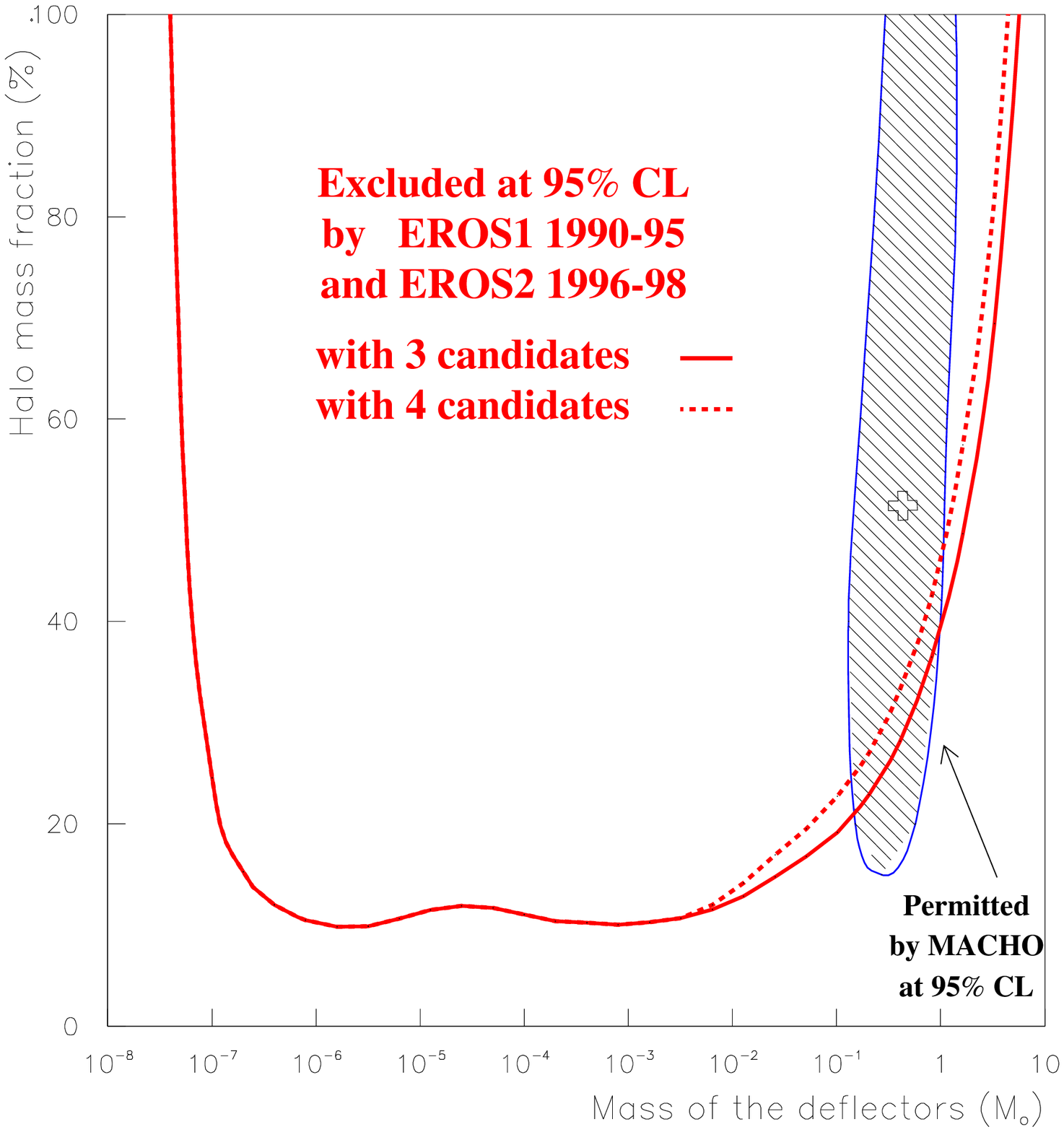,height=7cm}
\epsfig{file=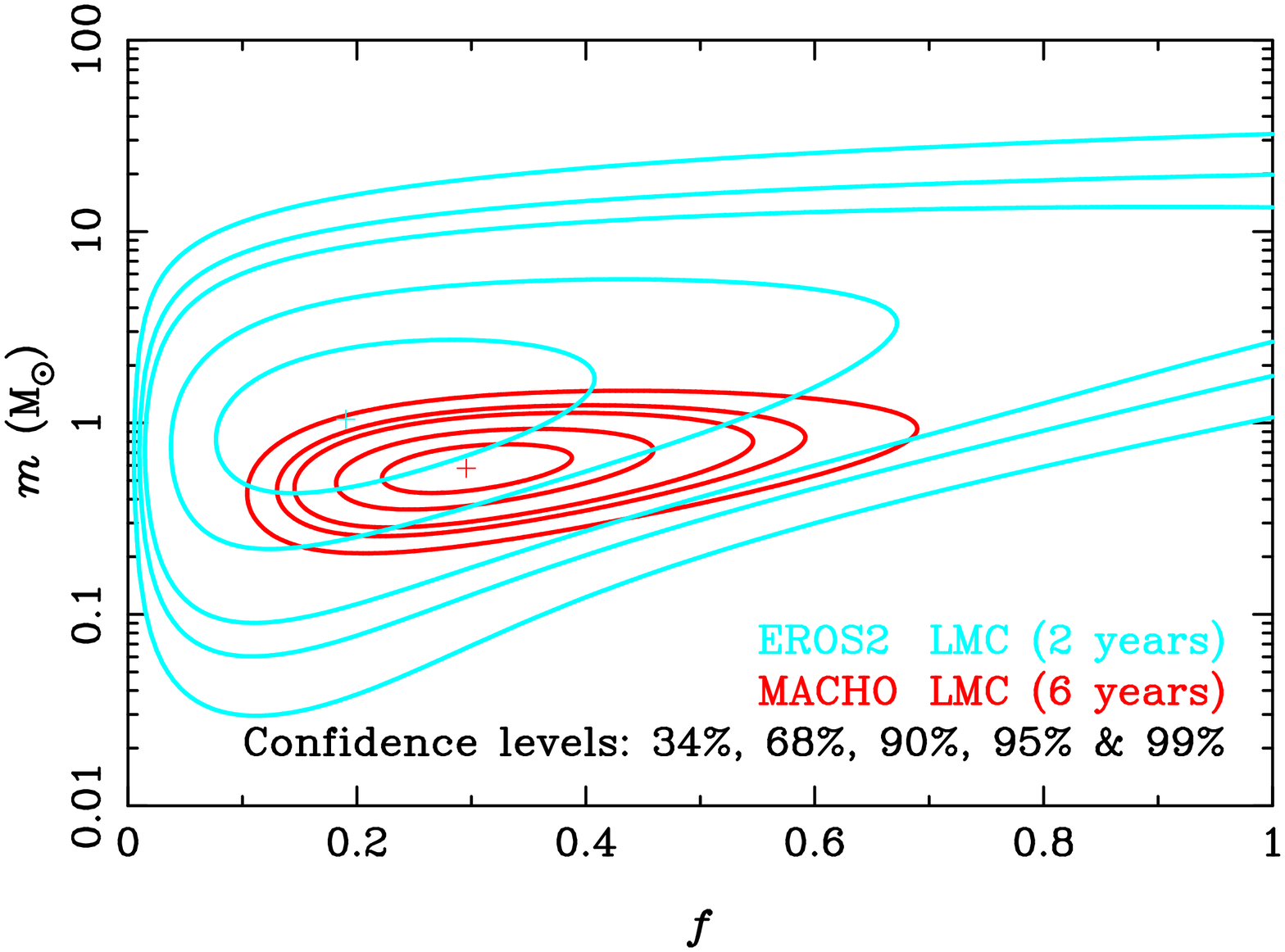,height=6cm}
\caption{{\em Left panel}\/: Upper limits on the halo fraction from
the \eros1+2 \ LMC and SMC experiments, with the \macho \ two-year
LMC results shown by the shaded region (figure courtesy of \eros). {\em
Right panel}\/: Likelihood analysis of the MACHO fraction $f$ and mass
$m$ from the \eros2 two-year and \macho \ six-year LMC datasets,
assuming the detected events are due to MACHOs. Both figures are for a
``standard'' isothermal dark halo.}
\label{e-ml}
\end{figure}

\subsection{What we've learned so far}

Before discussing the different interpretations of the microlensing
results we should first reflect upon several conclusions which we are
led to from both the \eros \ and \macho \ datasets, and which have
important implications for our understanding of Galactic dark matter.
They underscore the considerable success of microlensing so
far.

It is important to emphasize that the \macho \ and \eros \ LMC/SMC
results are statistically consistent with one another, as can be seen
in Figure~\ref{e-ml}. The absence of short-duration events in either
survey limits the contribution of low-mass MACHOs \cite{alc98}. The
\eros1+2 LMC and SMC limits in Figure~\ref{e-ml} indicate that $f <
0.12$ for MACHOs in the mass range $10^{-6} - 10^{-2}~\sm$.  Both
\eros \ and \macho \ exclude brown dwarfs as a major constituent of
the dark matter and both also agree that MACHOs of around a Solar mass or
below do not dominate the halo dark matter budget. If one chooses to
interpret the detected events as being due to MACHOs, both \eros \ and
\macho \ LMC datasets prefer a MACHO fraction $f \sim 0.2-0.3$ and
mass $m \sim 0.5-1~\sm$. However, the teams themselves have
interpreted their results in contrasting fashion, with \eros \
choosing to place only upper limits on $f$ (as shown in the left panel
in Figure~\ref{e-ml}) and \macho \ arguing that its dataset indicates
a positive detection of MACHOs (as shown in the right-hand panel). The
uncertainty is not whether microlensing has truly been observed, but
whether it being caused by MACHOs.

\section{MACHO solutions}

\subsection{Dependency of MACHO mass on halo model}

Analyses of the MACHO mass inferred from microlensing data always
assume some underlying model for the distribution function of the
Galactic halo.  The structure of the Galactic halo is very uncertain,
so one might expect a large systematic error in MACHO mass
determinations. Various studies employing a range of flattened halo
models and anisotropic velocity distributions show that the MACHO mass
determination is actually rather robust. In particular, brown dwarfs
appear to be irreconcilable with current datasets \cite{gates1}. If
the microlensing events are due to MACHOs then their mass is of the
order of a Solar mass.

\begin{figure}
\begin{center}
\epsfig{file=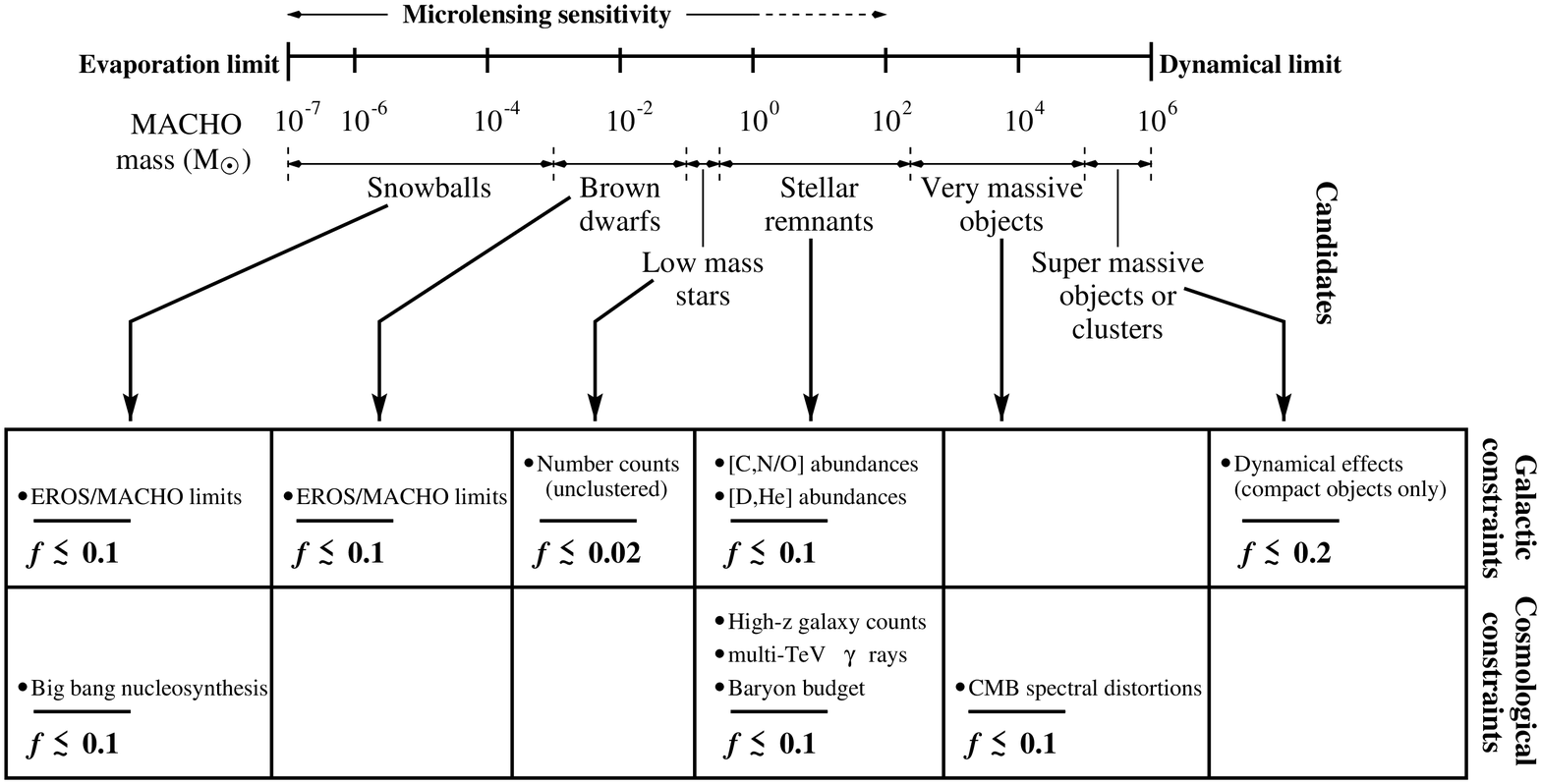,width=14cm}
\end{center}
\caption{Baryonic MACHO candidates and constraints}
\label{const}
\end{figure}

\subsection{Baryonic MACHOs}

Figure~\ref{const} shows the range of baryonic MACHO candidates, as
well as a summary of present constraints on them \cite{carr94}. The
microlensing mass sensitivity currently spans about seven of the
thirteen orders of magnitude of the candidates, and provides the
strongest constraints on snowballs and brown dwarfs. Other constraints
involve a wide range of astrophysical arguments but can be loosely
classified into those derived from properties of our Galaxy and those
derived from cosmological considerations. In order to translate the
latter into limits on the halo fraction $f$ for our own Galaxy, one must
assume it to be cosmologically representative, which may or may not be
the case. 

It is interesting to note that the combination of Galactic and
cosmological constraints now limits the abundance of all the baryonic
MACHO candidates to well below that required to explain the halo dark
matter. In fact, even if we populate the halo with a mixture of
baryonic candidates, they can provide only just over half of the dark
matter if they are to remain consistent with all constraints. However, 
low-mass stars are the only candidates to be excluded
by direct observation.
The candidate which is most compatible with the microlensing mass
scale is the white dwarf. Ibata et al. have detected a number of faint
moving objects in the Hubble Deep Field (HDF) consistent with a
population of white dwarfs contributing a significant fraction of the
halo dark matter \cite{iba1}. Subsequently, Ibata et al. obtained
spectroscopy of two nearby high proper motion white dwarfs, inferring
these to be possibly the nearby counterparts of the HDF objects. The
observations clearly lend support to the microlensing results.

These positive findings contrast with the range of constraints in
Figure~\ref{const}.  Remnants are constrained by both
Galactic and cosmological arguments, which all indicate $f \la 0.1$,
barely allowing consistency with the microlensing results \cite{Free}.
Galactic limits relate to the production
of helium and metals, and the depletion of deuterium by the precursor
stars. Cosmological limits come from the lack of precursor
starlight at high-redshift, the observation that the Universe appears
to be optically thin to TeV $\gamma$ rays out to the redshift of
blazar Mkn~501 at $z = 0.034$ (hence the infrared background
indicative of the precursor stars is low or not present), and
from the fact that too much of the ``baryon budget'' allowed by
Big-Bang nucleosynthesis predictions may still be in gas by $z \sim
1$. All of these arguments can be countered on an individual
basis. For example, the recent positive detection of an infrared
background \cite{wright} casts doubt on our understanding of the intrinsic
spectrum of Mkn~501, whilst the latest cosmic microwave background
data from {\sc Boomerang} favour a baryon density $50\%$ larger than
nucleosynthesis models predict \cite{boom}. However, the fact that these
diverse constraints are all consistent with each other seriously
undermines the white-dwarf hypothesis.

A recently proposed alternative to the white-dwarf solution is the
beige dwarf, a kind of ``genetically-modified'' brown dwarf. Hansen
\cite{hansen} has shown that slow accretion of gas onto a brown dwarf
can prevent the core temperature from rising to the point where
hydrogen-burning commences. Beige dwarfs can have masses up to
$0.3~\sm$ if they continue to accrete at the maximum rate over a
Hubble time, so providing compatibility with microlensing data.

\subsection{Non-baryonic MACHOs}

Though MACHOs are generally assumed to be baryonic there are a few
non-baryonic cold dark matter (CDM) candidates, such as primordial
black holes (PBHs), which would give rise to microlensing
signatures. The fact that microlensing searches now exclude more than
half of the dark matter from comprising MACHOs may be both good news
and bad for CDM. On the one hand it supports the view that at least
half of the dark matter is non-baryonic (though, conceivably, it could
also comprise cold clumps of gas). On the other hand it limits the
contribution of non-baryonic MACHO candidates as much as it does
baryonic dark matter. Furthermore, if the detected microlensing events
are, at least in part, due to MACHOs then we need either a combination
of baryonic dark matter and CDM or {\em two}\/ types of CDM: one to
provide the MACHOs and one to provide the rest of the dark
matter. This constitutes at least an aesthetic constraint: if MACHOs
are being detected it may be easier to believe they are baryonic than
to believe in two species of CDM.

The PBH scenario also requires fine-tuning. A first-order phase
transition at the QCD epoch provides a natural formation mechanism to
produce PBHs of about the right mass \cite{jed97}, however the PBH
abundance must be finely tuned in order that they do not rapidly
dominate the energy density of the early Universe as it expands.

\section{Non-MACHO solutions}

Do the microlensing results require MACHOs at all? We now examine the
arguments for and against a number of alternative solutions.

\subsection{Milky Way disk}

The microlensing contribution of a standard Milky Way disk is an order
of magnitude too small to account for the LMC events. Evans et
al. \cite{evans} investigated the possibility that the disk may be
significantly flared and warped in the direction of the Magellanic
Clouds. Whilst such a model could account for the events, a subsequent
study ruled out this proposal on the basis of star
counts \cite{evans}. Gates \& Gyuk \cite{gates2} have advocated the
existence of an old, super-thick disk with a scale height of $\sim
3$~kpc comprising white-dwarf remnants. The model succeeds in evading
the constraints on white dwarfs because their total mass is much less
than required by halo models. Whilst the model is strictly a very fat
disk, it can also be viewed as a strongly dissipated MACHO halo. It is
an {\em ad hoc}\/ solution but, more importantly, a currently viable
one.

\subsection{Intervening debris}

Zhao \cite{zhao} has considered the effects of stellar ``debris''
along the line of sight to the Magellanic Clouds, either associated in
some way with the Clouds or simply a chance alignment of some
disrupted satellite galaxy. The proposal received tentative
observational support from Zaritsky \& Lin \cite{zar}, who interpreted
an observed vertical HR extension to the LMC red clump population as
evidence of a foreground stellar population. Beaulieu \& Sackett \cite{sack}
claimed that such a feature is indicative of stellar evolution rather
than a foreground population, whilst Bennett argued that if the
structure has a similar star formation history to the LMC its optical
depth would be an order of magnitude too small to explain the \macho \
results \cite{sack}. Gould \cite{gou} has presented a series of
arguments against intervening populations. If they are unassociated
with the Magellanic Clouds then they must have a highly improbable
spatial and velocity alignment with the Clouds to have remained
undetected. On the other hand, if the debris is associated with the
Clouds, Gould argues that its mass function would need to be dominated
by sub-stellar objects in order to explain the microlensing
timescales. However, it has been suggested recently that such a
population may have been observed \cite{graf}. Of course, it is
possible that the debris is mostly dark, in which case it is
essentially a MACHO solution in disguise.

\subsection{LMC/SMC self-lensing}

The last possibility is that the sources in the Magellanic Clouds
themselves are also providing the lenses \cite{sahu}. Gould
\cite{gou2} has shown that if the LMC can be represented by a thin
disk in virial equilibrium then its optical depth would be about an
order of magnitude too small to explain the microlensing
results. Though this may be too strict an assumption for what is a
poorly understood structure, recent self-consistent numerical models yield
similar optical depths \cite{lmchalo}.

The strongest support for the self-lensing scenario may come from the
microlensing data. The two binary caustic crossing events, MACHO LMC-9
and 98-SMC-1, provide information on their line-of-sight
location. Their caustic-crossing timescales indicate the time taken
for the caustic to cross the face of the source and so depends on the
size of the star and the projected transverse velocity of the
lens. The typical transverse velocity of halo or Galactic disk lenses
is of the order of 1000~km~s$^{-1}$ or more, where as for lenses in
the Clouds it is typically only $\sim 60-80$~km~s$^{-1}$. In the case
of 98-SMC-1, the second caustic crossing is well resolved \cite{alc00}
and the projected velocity is determined to be 65--75~km~s$^{-1}$,
consistent with it being within the SMC and highly inconsistent with
it being a MACHO. The second caustic crossing of MACHO~LMC-9 is only
partially resolved \cite{ben96} and the projected velocity is inferred
to be only 20~km~s$^{-1}$, which is low even for a lens within the
LMC, but certainly excludes a MACHO interpretation. Statistically, the
two binary events strongly favour self-lensing over Galactic MACHOs
\cite{ker1}. However, it is possible that the A-type source of event
LMC-9 may itself be an equal luminosity binary \cite{ben96}, in which
case the timescale is indicative of the binary separation and the
projected transverse velocity could be large enough to be consistent
with the lens being a MACHO after all.

Finally, the lenses could reside in a dark halo
associated with the Clouds, rather than the Milky Way halo
\cite{ker1}. Alves \& Nelson \cite{lmchalo} have recently argued that
the LMC rotation curve does not require a halo, though in any case an
LMC halo must count as a MACHO solution.

\section{Summary and future prospects}

Despite the many successes of microlensing, as yet there is no firm
evidence that we are detecting MACHOs. Of the proposed solutions
involving MACHOs, halo white dwarfs appear to be all but excluded by a
host of constraints, whilst beige dwarfs and primordial black holes,
though not constrained, require finely-tuned formation scenarios. As
far as non-MACHO solutions are concerned, visible stellar populations
do not appear to be able to provide the observed optical depth. Viable
models include a dark super-thick disk, mostly dark tidal debris, or
dark haloes around the Magellanic Clouds. Each of these solutions
requires a major revision in our understanding of the structure of our
Galaxy or its satellites, and each is really a MACHO solution in a new
form.

In order to resolve the issue what is required is the ability to
identify MACHO events from other microlensing events. The Andromeda
galaxy presents an attractive possibility. If it is surrounded by a
spherical halo of MACHOs the microlensing rate to the far side of its
inclined disk should be larger than the rate towards the near
side. The detection of such a gradient would provide clear evidence of
MACHO rather than stellar lensing. Experiments are now in progress to
try to detect this signal \cite{ker2}. For our own Galaxy, the spatial
distribution of the LMC microlensing events, or the velocity or
magnitude distributions of the sources \cite{zhao2}, may soon provide
a conclusive answer. If not, measurements by astrometric satellites
such as SIM \cite{gou3} or GAIA could settle the issue by determining the
line-of-sight location of a handful of events.

\section*{Acknowledgments}
I am grateful to the organisers and the EU TMR programme for partial
funding to attend this meeting, and to PPARC for supporting
my postdoctoral fellowship.

\end{document}